\documentstyle[12pt]{article}
\def\newpic#1{%
   \def\emline##1##2##3##4##5##6{%
      \put(##1,##2){\special{em:point #1##3}}%
      \put(##4,##5){\special{em:point #1##6}}%
      \special{em:line #1##3,#1##6}}}
\begin{document}
\begin{titlepage}
{}~
\begin{flushright}
Preprint\\
ITP-95-29E
\end{flushright}
\vskip 4.cm
\begin{center}
{\Large \bf Unitarized model\\
\vskip .2 cm
of hadronic diffractive
dissociation}
\vskip 1.cm
{\large E.S.Martynov, B.V.Struminsky}
\vskip 1.5cm
N.N.Bogoliubov Institute for Theoretical Physics,\\
252143, Metrologicheskaja st. 14b, Kiev-143, Ukraine\\
\vskip 2.cm
{\large Abstract}
\end{center}
\vskip .5cm
It is shown that in a supercritical pomeron model the contribution of the
tirple-pomeron diagrams violates the unitarity bound for cross-section
even with account of the multiple pomeron exchanges between the initial
hadrons. Asymptotic behaviour of the single diffractive dissoci\-ation
cross-section is calculated in the approximation where every
pomeron in the $3P$-diagram is eikonalized as well as an elastic interaction
of initial hadrons is taken into account. In this approximation
$\sigma^{SD}/\sigma_{tot}\rightarrow 0$ at $s\rightarrow \infty.$
\end{titlepage}

The interest in diffractive dissociation is caused by both
comparatively new experimental data obtained at the Tevatron [1] and by
observation at HERA of deep inelastic processes with the typical
rapidity gaps [2]. Generally adopted interpretation of these similar enough
phenomena is based on the following interaction mechanism. The
incoming fast proton (or virtual photon in deep inelastic scattering)
"emits" a pomeron which interacts with the proton-target
producing a shower of hadrons. These hadrons are distributed in the
rapidity scale at a large distance (rapidity gap) from the initial proton.
The central point of the model, the interaction of pomeron with
proton-target, is universal. It means that this subprocess does not
depend on where a dissociation is considered: in a pure hadronic process or
in a deep inelastic scattering.

In what follows we consider diffractive dissociation in a pure
hadronic process $hh\rightarrow hX$. If the effective mass of produced
shower is large
enough then with certain simplifying assumptions the cross-section
of the process may be presented (due to the generalized optical theorem)
by the diagram with a triple-pomeron vertex (Fig.1)
\vskip 1.5cm
\begin{center}
\newpic{}
\hskip -5.cm
\unitlength=0.60mm
\special{em:linewidth 0.6pt}
\linethickness{0.6pt}
\begin{picture}(210.95,139.00)
\put(75.43,100.00){\circle{14.00}}
\emline{47.43}{100.00}{1}{68.43}{100.00}{2}
\emline{68.43}{100.00}{3}{68.43}{100.00}{4}
\emline{81.43}{104.00}{5}{107.43}{104.00}{6}
\emline{82.43}{100.00}{7}{107.43}{100.00}{8}
\emline{81.43}{96.00}{9}{107.43}{96.00}{10}
\emline{107.43}{96.00}{11}{107.43}{96.00}{12}
\emline{47.43}{136.00}{13}{107.43}{136.00}{14}
\emline{75.43}{136.00}{15}{81.43}{130.00}{16}
\emline{81.43}{130.00}{17}{71.43}{126.00}{18}
\emline{71.43}{126.00}{19}{81.43}{120.00}{20}
\emline{81.43}{120.00}{21}{71.43}{116.00}{22}
\emline{71.43}{116.00}{23}{81.43}{109.00}{24}
\emline{81.43}{109.00}{25}{74.43}{107.00}{26}
\emline{44.43}{139.00}{27}{44.43}{90.00}{28}
\emline{120.05}{139.00}{29}{120.05}{90.00}{30}
\put(130.81,114.00){\makebox(0,0)[cc]{=}}
\put(123.05,137.00){\makebox(0,0)[cc]{2}}
\emline{144.95}{90.00}{31}{210.95}{90.00}{32}
\emline{144.95}{124.00}{33}{161.95}{124.00}{34}
\emline{210.95}{124.00}{35}{192.95}{124.00}{36}
\emline{161.95}{124.00}{37}{168.95}{137.00}{38}
\emline{192.95}{124.00}{39}{185.95}{137.00}{40}
\emline{177.95}{109.00}{41}{175.95}{111.00}{42}
\emline{175.95}{111.00}{43}{175.95}{114.00}{44}
\emline{175.95}{114.00}{45}{171.95}{114.00}{46}
\emline{171.95}{114.00}{47}{171.95}{118.00}{48}
\emline{171.95}{118.00}{49}{167.95}{118.00}{50}
\emline{167.95}{118.00}{51}{167.95}{122.00}{52}
\emline{167.95}{122.00}{53}{163.95}{122.00}{54}
\emline{163.95}{122.00}{55}{161.95}{124.00}{56}
\emline{177.95}{109.00}{57}{179.95}{111.00}{58}
\emline{179.95}{111.00}{59}{179.95}{114.00}{60}
\emline{179.95}{114.00}{61}{183.95}{114.00}{62}
\emline{183.95}{114.00}{63}{183.95}{118.00}{64}
\emline{183.95}{118.00}{65}{187.95}{118.00}{66}
\emline{187.95}{118.00}{67}{187.95}{122.00}{68}
\emline{187.95}{122.00}{69}{191.95}{122.00}{70}
\emline{191.95}{122.00}{71}{192.95}{124.00}{72}
\emline{177.95}{109.00}{73}{177.95}{106.00}{74}
\emline{177.95}{106.00}{75}{175.95}{104.00}{76}
\emline{175.95}{104.00}{77}{179.95}{100.00}{78}
\emline{179.95}{100.00}{79}{175.95}{96.00}{80}
\emline{175.95}{96.00}{81}{179.95}{92.00}{82}
\emline{179.95}{92.00}{83}{176.95}{90.00}{84}
\put(163.95,112.00){\makebox(0,0)[cc]{P}}
\put(190.95,112.00){\makebox(0,0)[cc]{P}}
\put(170.95,100.00){\makebox(0,0)[cc]{P}}
\put(83.43,117.00){\makebox(0,0)[cc]{P}}
\put(111.90,99.05){\makebox(0,0)[cc]{X}}
\put(39.05,114.76){\makebox(0,0)[cc]{${\bf \Sigma}_{X}^{}$}}
\put(104.00,75.00){\makebox(0,0)[cc]{Fig.1 Process of difractive dissociation
and 3P-diagramm}}
\end{picture}
\end{center}
\vskip -4.4cm
The properties of the cross-section of course depend on specific model
of pomeron. The most known one at present, the so-called supercritical pomeron
has a trajectory with the intercept $\alpha_{P}(0)=1+\Delta $ with
$\Delta >0$. In particular the model of A.Donnachie and P.Landshoff
with $\Delta=0.08$ based on a pomeron-photon analogy describes quite
well the hadronic data [3]. However the contribution of supercritical
pomeron to the total cross-section rises with energy as a power $\sigma
\propto s^{\Delta}$, violating the Froissart-Martin bound
$\sigma_{tot}<const \ln^{2}(s/s_{0})$. The strict consistent procedure of
unitarization is absent now, but there are some simple phenomenological
ways to eliminate the rough contradictions with the unitarity. For
example, the eikonal [4], $U$-matrix [5] methods and their
generalizations are often used for unitarization of the amplitude.
Another approach to the problem was suggested in [6].

It is quite obvious that the three-pomeron diagram needs also for
unitarity corrections, which should remove a too fast growing
contribution of supercritical pomeron to the diffractive
dissociation cross-section (up to the $\ln s$-factors it is
proportional to $s^{2\Delta}$). The $3P$-diagram seemed to be
unitarized by the most simple way, taking into account multiple
pomeron exchanges between the incoming hadrons (initial state
interaction). This approach was suggested in the Ref.[7]. In a more
general content a problem of unitarization in diffractive dissociation
was discussed a long time ago (see for example [8]). It was
claimed in [7] that with the account of the initial state interaction,
the integrated cross-section of diffractive dissociation rises
logarithmicaly with energy, $\sigma^{SD}\propto \ln s$, in accordance
with the experimental data.

We will show that this conclusion of [7] is wrong, and the account of
initial
state interaction by the eikonal (or by another) way does not allow to
eliminate too fast growth of cross-section. We have performed the
asymptotic evaluations of more wide class of corrections, which indeed
allow to eliminate the explicit contradiction with unitarity bounds.

To make our arguments more clear we list a few well known general
statements and formulae. Because we are interested only in an asymptotic
cross-section behaviour, a contribution of $f$-reggeon is omitted in all
expressions.

As in [7] we will work in the impact parameter representation. The
normalization of amplitude is
$$\frac{d\sigma}{dt} =
\pi|f(s,t)|^{2},\qquad \sigma_{tot} = 4\pi Imf(s,0).\eqno(1)$$
An amplitude in $b$-representation is defined by the
following transformation
$$a(s,b) =
\frac{1}{2\pi}\int d{\vec q}e^{-i{\vec q}\,{\vec b}}f(s,t),\quad
t=-q^{2}.\eqno(2)$$

Eikonal summation of the high energy elastic pomeron rescatterings
can be realized with the amplitude $a(s,b)$ in the form
$$a(s,b) = i(1-e^{-\Omega(s,b)}),\qquad \Omega(s,b) =
-\frac{i}{2\pi}\int d{\vec q}e^{-i{\vec q}\,{\vec b}}f_{0}(s,t),$$
where $f_{0}(s,t)$ is an input elastic amplitude.
Starting from a simplified model of supercritical pomeron with the
trajectory $\alpha_{P}(t)=1+\Delta+\alpha'_{P}t$
$$f_{0}(s,t) =
ig(t)\biggl (\frac{s}{s_{0}}\biggr)^{\alpha_{P}(t)-1},\qquad g(t) =
g\,e^{-B_{0}t/4},$$
one can obtain
$$\Omega(s,b)=\nu(s/s_{0})e^{-b^{2}/R^{2}(s/s_{0})},\eqno(3)$$
where
$$\nu(s/s_{0}) = \frac{\sigma_{0}}{2\pi
R^{2}(s/s_{0})}\biggl(\frac{s}{s_{0}}\biggl)^{\Delta}=
\frac{4g^{2}}{R^{2}(s/s_{0})}
\biggl(\frac{s}{s_{0}}\biggl)^{\Delta},\eqno(4)$$
$$R^{2}(s/s_{0}) = 2B_{0} +4 \alpha'\ln (s/s_{0}),\qquad
\sigma_{0}=8\pi g^{2}.\eqno(5)$$
In this model $if_{0}(s,t)$ and $\Omega(s,b)$ are the real functions,
but analiticity and crossing-symmetry  are restored by the substitution
$s\rightarrow s\exp(-i\pi /2)$. It is easy to obtain from the above
expressions that
$\sigma_{tot} \simeq 2\pi \Delta R^{2}(s/s_{0})\ln(s/s_{0})$
at $s/s_{0}\rightarrow \infty $
Thus, in a supercritical pomeron model the eikonal corrections to
one-pomeron exchange remove the explicit violation of unitarity
condition. The resulting cross-sections satisfies the Froissart-Martin
bound.

The expression for an integrated over $t$ cross-section of
diffractive dissociation was written in [7] in the form
$$
M^{2}\frac{d\sigma^{SD}}{dM^{2}}  =
\sigma_{0}^{2}G_{PPP} \biggl(\frac{s}{M^{2}}\biggr)^{2\Delta}
\biggl(\frac{M^{2}}{s_{0}}\biggr)^{\Delta}
\frac{1}{[\pi \tilde R^{2}(s/M^{2})]^{2}\pi \tilde R^{2}(M^{2}/s_{0})}$$
$$\times \int d{\vec b}\,d{\vec b'}
\exp\biggl(-2\nu(s/s_{0})e^{-b^{2}/R^{2}(s/s_{0})}\biggr)
\exp\biggl(-2\frac{({\vec
b}-{\vec{b'}})^{2}}{\tilde R^{2}(s/M^{2})}-\frac{b'^{2}}{\tilde
R^{2}(M^{2}/s_{0})}\biggr),\eqno(6)$$
where $\nu, R^{2}$ are defined by exps.(4),(5),
$$\tilde R^{2}(z) = B_{0} + r^{2} + 4\alpha'\ln (z)\eqno(7)$$
and $r$ is radius of the triple-pomeron vertex.
The eikonal corrections due to pomeron rescatterings in initial
state (Fig.2) were accounted by the insertion of factor $
\exp(-2\Omega(s,b))=
\exp\biggl(-2\nu(s/s_{0})e^{-b^{2}/R^{2}(s/s_{0})}\biggr)$ in the
integrand of the exp.(6)
\begin{center}
\newpic{}
\unitlength=0.80mm
\special{em:linewidth 0.6pt}
\linethickness{0.4pt}
\begin{picture}(115.00,70.00)
\emline{25.95}{23.00}{1}{91.95}{23.00}{2}
\emline{25.95}{57.00}{3}{42.95}{57.00}{4}
\emline{91.95}{57.00}{5}{73.95}{57.00}{6}
\emline{42.95}{57.00}{7}{49.95}{70.00}{8}
\emline{73.95}{57.00}{9}{66.95}{70.00}{10}
\emline{58.95}{42.00}{11}{56.95}{44.00}{12}
\emline{56.95}{44.00}{13}{56.95}{47.00}{14}
\emline{56.95}{47.00}{15}{52.95}{47.00}{16}
\emline{52.95}{47.00}{17}{52.95}{51.00}{18}
\emline{52.95}{51.00}{19}{48.95}{51.00}{20}
\emline{48.95}{51.00}{21}{48.95}{55.00}{22}
\emline{48.95}{55.00}{23}{44.95}{55.00}{24}
\emline{44.95}{55.00}{25}{42.95}{57.00}{26}
\emline{58.95}{42.00}{27}{60.95}{44.00}{28}
\emline{60.95}{44.00}{29}{60.95}{47.00}{30}
\emline{60.95}{47.00}{31}{64.95}{47.00}{32}
\emline{64.95}{47.00}{33}{64.95}{51.00}{34}
\emline{64.95}{51.00}{35}{68.95}{51.00}{36}
\emline{68.95}{51.00}{37}{68.95}{55.00}{38}
\emline{68.95}{55.00}{39}{72.95}{55.00}{40}
\emline{72.95}{55.00}{41}{73.95}{57.00}{42}
\emline{58.95}{42.00}{43}{58.95}{39.00}{44}
\emline{58.95}{39.00}{45}{56.95}{37.00}{46}
\emline{56.95}{37.00}{47}{60.95}{33.00}{48}
\emline{60.95}{33.00}{49}{56.95}{29.00}{50}
\emline{56.95}{29.00}{51}{60.95}{25.00}{52}
\emline{60.95}{25.00}{53}{57.95}{23.00}{54}
\emline{27.00}{57.00}{55}{31.00}{53.00}{56}
\emline{31.00}{53.00}{57}{27.00}{49.00}{58}
\emline{27.00}{49.00}{59}{31.00}{45.00}{60}
\emline{31.00}{45.00}{61}{27.00}{41.00}{62}
\emline{27.00}{41.00}{63}{31.00}{37.00}{64}
\emline{31.00}{37.00}{65}{27.00}{33.00}{66}
\emline{27.00}{33.00}{67}{31.00}{29.00}{68}
\emline{31.00}{29.00}{69}{27.00}{25.00}{70}
\emline{27.00}{25.00}{71}{29.00}{23.00}{72}
\emline{29.00}{23.00}{73}{29.00}{23.00}{74}
\emline{32.00}{57.00}{75}{36.00}{53.00}{76}
\emline{36.00}{53.00}{77}{32.00}{49.00}{78}
\emline{32.00}{49.00}{79}{36.00}{45.00}{80}
\emline{36.00}{45.00}{81}{32.00}{41.00}{82}
\emline{32.00}{41.00}{83}{36.00}{37.00}{84}
\emline{36.00}{37.00}{85}{32.00}{33.00}{86}
\emline{32.00}{33.00}{87}{36.00}{29.00}{88}
\emline{36.00}{29.00}{89}{32.00}{25.00}{90}
\emline{32.00}{25.00}{91}{34.00}{23.00}{92}
\emline{34.00}{23.00}{93}{34.00}{23.00}{94}
\emline{26.00}{23.00}{95}{17.00}{23.00}{96}
\emline{26.00}{57.00}{97}{17.00}{57.00}{98}
\emline{92.00}{57.00}{99}{103.00}{57.00}{100}
\emline{92.00}{23.00}{101}{103.00}{23.00}{102}
\emline{21.00}{57.00}{103}{25.00}{53.00}{104}
\emline{25.00}{53.00}{105}{21.00}{49.00}{106}
\emline{21.00}{49.00}{107}{25.00}{45.00}{108}
\emline{25.00}{45.00}{109}{21.00}{41.00}{110}
\emline{21.00}{41.00}{111}{25.00}{37.00}{112}
\emline{25.00}{37.00}{113}{21.00}{33.00}{114}
\emline{21.00}{33.00}{115}{25.00}{29.00}{116}
\emline{25.00}{29.00}{117}{21.00}{25.00}{118}
\emline{21.00}{25.00}{119}{23.00}{23.00}{120}
\emline{23.00}{23.00}{121}{23.00}{23.00}{122}
\emline{84.00}{57.00}{123}{88.00}{53.00}{124}
\emline{88.00}{53.00}{125}{84.00}{49.00}{126}
\emline{84.00}{49.00}{127}{88.00}{45.00}{128}
\emline{88.00}{45.00}{129}{84.00}{41.00}{130}
\emline{84.00}{41.00}{131}{88.00}{37.00}{132}
\emline{88.00}{37.00}{133}{84.00}{33.00}{134}
\emline{84.00}{33.00}{135}{88.00}{29.00}{136}
\emline{88.00}{29.00}{137}{84.00}{25.00}{138}
\emline{84.00}{25.00}{139}{86.00}{23.00}{140}
\emline{86.00}{23.00}{141}{86.00}{23.00}{142}
\emline{90.00}{57.00}{143}{94.00}{53.00}{144}
\emline{94.00}{53.00}{145}{90.00}{49.00}{146}
\emline{90.00}{49.00}{147}{94.00}{45.00}{148}
\emline{94.00}{45.00}{149}{90.00}{41.00}{150}
\emline{90.00}{41.00}{151}{94.00}{37.00}{152}
\emline{94.00}{37.00}{153}{90.00}{33.00}{154}
\emline{90.00}{33.00}{155}{94.00}{29.00}{156}
\emline{94.00}{29.00}{157}{90.00}{25.00}{158}
\emline{90.00}{25.00}{159}{92.00}{23.00}{160}
\emline{92.00}{23.00}{161}{92.00}{23.00}{162}
\emline{96.00}{57.00}{163}{100.00}{53.00}{164}
\emline{100.00}{53.00}{165}{96.00}{49.00}{166}
\emline{96.00}{49.00}{167}{100.00}{45.00}{168}
\emline{100.00}{45.00}{169}{96.00}{41.00}{170}
\emline{96.00}{41.00}{171}{100.00}{37.00}{172}
\emline{100.00}{37.00}{173}{96.00}{33.00}{174}
\emline{96.00}{33.00}{175}{100.00}{29.00}{176}
\emline{100.00}{29.00}{177}{96.00}{25.00}{178}
\emline{96.00}{25.00}{179}{98.00}{23.00}{180}
\emline{98.00}{23.00}{181}{98.00}{23.00}{182}
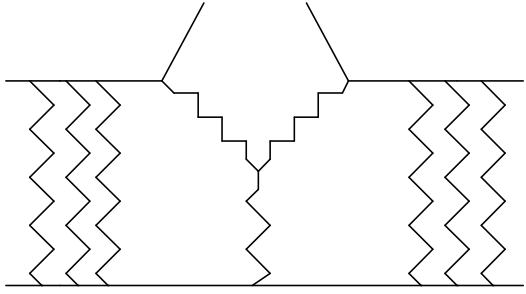
\put(45.00,04.00){\makebox(0,0)[cc]{Fig. 2. 3P-diagram with the
interaction of hadrons in the initial state}}
\end{picture}
\end{center}

Unfortunately a mistake was appeared in [7] when asymptotic evaluation
of the integral was made. After integration over $b$ and $b'$
the differential cross-section of diffractive
dissociation becomes as following
$$\frac{M^{2}d\sigma^{SD}}{dM^{2}}=\frac{\sigma_{0}^{2}}{2\pi \tilde
R^{2}_{1}(s/M^2)} G_{PPP}\frac{a\gamma[a,2\nu(s)]}
{[2\nu(s)]^{a}}\biggl(\frac{s}{M^{2}}\biggr)^{2\Delta}
\biggl(\frac{M^{2}}{s_{0}}\biggr)^{\Delta},\eqno(8)$$
where
$$a = \frac{2R^{2}(s/s_{0})}{\tilde R^{2}(s/M^2) + 2\tilde
R^{2}(M^2/s_{0})},\eqno(9)$$
and $\gamma[a,2\nu]$ is the incomplete Euler gamma function.

In the limit under consideration,
$s\gg s_{0}, M^{2}/s_{0}, s/M^{2}\gg 1,$ the ratio $a$ tends to $2$
and $\gamma[a,2\nu]$ tends to $\Gamma(2)$. Substituting these limits
to the expression (8) authors of [7] have obtained
$$\frac{M^{2}d\sigma^{SD}}{dM^{2}} = \pi R^{2}(s/s_{0})G_{PPP}
\biggl(\frac{M^{2}}{s_{0}}\biggr)^{-\Delta}.$$

However, this result is wrong. Indeed, one can see
using the definitions (4),(5),(7) and (9) that at
$s,\,M^{2},\,s/M^{2}\rightarrow \infty$
$$a = \frac{2R^{2}(s/s_{0})}{\tilde R^{2}_{1}(s/M^2) + 2\tilde
R^{2}_{1}(M^2/s_{0})} = 2\biggl(1 - \frac{\ln (M^{2}/s_{0})}{\ln
(s/s_{0})} + o(\frac{1}{\ln (s/s_{0})})\biggr),$$
Therefore the factor of expression (9) which violates the unitarity is
transformed as follows
$$ \biggl(\frac{s}{M^{2}}\biggr)^{2\Delta}\biggl(\frac{M^{2}}{s_{0}}\biggr)^
{\Delta} [\nu(s)]^{-a} \simeq
\exp\biggl
\{2\Delta\ln(\frac{s}{M^{2}})+\Delta\ln(\frac{M^{2}}{s_{0}})-
2\Delta\ln(s/s_{0})]\biggr\}\sim
\biggl(\frac{M^{2}}{s_{0}}\biggr)^{\Delta},$$
conserving too fast growth of cross-section at large $M^2$.

In what follows we investigate the diagrams which are
important to restore the unitarity. In our opinion it is necessary
not only to take into account an interaction of hadrons in the initial
state but
also to "eikonalize" each pomeron in the diagram of Fig.2. In
another words we estimate an asymptotical contribution to
$\sigma^{SD}$ of diagrams of Fig.3.
\begin{center}
\newpic{}
\unitlength=0.60mm \special{em:linewidth 0.6pt}
\linethickness{0.4pt}
\begin{picture}(178.24,97.00)
\put(90.10,45.00){\circle{14.00}}
\emline{91.10}{10.00}{1}{151.10}{10.00}{2}
\emline{151.10}{10.00}{3}{91.10}{10.00}{4}
\emline{151.10}{10.00}{5}{31.10}{10.00}{6}
\emline{31.10}{76.00}{7}{61.10}{76.00}{8}
\emline{121.10}{76.00}{9}{151.10}{76.00}{10}
\emline{94.10}{36.00}{11}{94.10}{36.00}{12}
\emline{94.10}{36.00}{13}{97.10}{33.00}{14}
\emline{97.10}{33.00}{15}{93.10}{29.00}{16}
\emline{93.10}{29.00}{17}{97.10}{25.00}{18}
\emline{97.10}{25.00}{19}{93.10}{21.00}{20}
\emline{93.10}{21.00}{21}{97.10}{17.00}{22}
\emline{97.10}{17.00}{23}{93.10}{13.00}{24}
\emline{93.10}{13.00}{25}{95.10}{11.00}{26}
\emline{94.10}{36.00}{27}{94.10}{36.00}{28}
\emline{94.10}{36.00}{29}{97.10}{33.00}{30}
\emline{97.10}{33.00}{31}{93.10}{29.00}{32}
\emline{93.10}{29.00}{33}{97.10}{25.00}{34}
\emline{97.10}{25.00}{35}{93.10}{21.00}{36}
\emline{93.10}{21.00}{37}{97.10}{17.00}{38}
\emline{97.10}{17.00}{39}{93.10}{13.00}{40}
\emline{93.10}{13.00}{41}{95.10}{11.00}{42}
\emline{86.10}{38.00}{43}{86.10}{36.00}{44}
\emline{86.10}{36.00}{45}{86.10}{36.00}{46}
\emline{86.10}{36.00}{47}{89.10}{33.00}{48}
\emline{89.10}{33.00}{49}{85.10}{29.00}{50}
\emline{85.10}{29.00}{51}{89.10}{25.00}{52}
\emline{89.10}{25.00}{53}{85.10}{21.00}{54}
\emline{85.10}{21.00}{55}{89.10}{17.00}{56}
\emline{89.10}{17.00}{57}{85.10}{13.00}{58}
\emline{85.10}{13.00}{59}{87.10}{11.00}{60}
\emline{86.10}{38.00}{61}{86.10}{36.00}{62}
\emline{86.10}{36.00}{63}{86.10}{36.00}{64}
\emline{86.10}{36.00}{65}{89.10}{33.00}{66}
\emline{89.10}{33.00}{67}{85.10}{29.00}{68}
\emline{85.10}{29.00}{69}{89.10}{25.00}{70}
\emline{89.10}{25.00}{71}{85.10}{21.00}{72}
\emline{85.10}{21.00}{73}{89.10}{17.00}{74}
\emline{89.10}{17.00}{75}{85.10}{13.00}{76}
\emline{85.10}{13.00}{77}{87.10}{11.00}{78}
\emline{86.10}{38.00}{79}{86.10}{39.00}{80}
\emline{94.10}{36.00}{81}{94.10}{37.00}{82}
\emline{94.10}{37.00}{83}{94.10}{39.00}{84}
\emline{61.10}{76.00}{85}{82.10}{97.00}{86}
\emline{121.10}{76.00}{87}{100.10}{97.00}{88}
\emline{58.10}{73.00}{89}{58.10}{70.00}{90}
\emline{58.10}{70.00}{91}{63.10}{70.00}{92}
\emline{63.10}{70.00}{93}{63.10}{65.00}{94}
\emline{63.10}{65.00}{95}{68.10}{65.00}{96}
\emline{68.10}{65.00}{97}{68.10}{60.00}{98}
\emline{68.10}{60.00}{99}{68.10}{60.00}{100}
\emline{68.10}{60.00}{101}{68.10}{59.00}{102}
\emline{68.10}{59.00}{103}{74.10}{59.00}{104}
\emline{74.10}{59.00}{105}{74.10}{53.00}{106}
\emline{74.10}{53.00}{107}{79.10}{53.00}{108}
\emline{79.10}{53.00}{109}{79.10}{47.00}{110}
\emline{79.10}{47.00}{111}{82.10}{47.00}{112}
\emline{60.10}{75.00}{113}{60.10}{72.00}{114}
\emline{60.10}{72.00}{115}{65.10}{72.00}{116}
\emline{65.10}{72.00}{117}{65.10}{67.00}{118}
\emline{65.10}{67.00}{119}{70.10}{67.00}{120}
\emline{70.10}{67.00}{121}{70.10}{62.00}{122}
\emline{70.10}{62.00}{123}{70.10}{62.00}{124}
\emline{70.10}{62.00}{125}{70.10}{61.00}{126}
\emline{70.10}{61.00}{127}{76.10}{61.00}{128}
\emline{76.10}{61.00}{129}{76.10}{55.00}{130}
\emline{76.10}{55.00}{131}{81.10}{55.00}{132}
\emline{81.10}{55.00}{133}{81.10}{49.00}{134}
\emline{81.10}{49.00}{135}{84.10}{49.00}{136}
\emline{62.10}{77.00}{137}{62.10}{74.00}{138}
\emline{62.10}{74.00}{139}{67.10}{74.00}{140}
\emline{67.10}{74.00}{141}{67.10}{69.00}{142}
\emline{67.10}{69.00}{143}{72.10}{69.00}{144}
\emline{72.10}{69.00}{145}{72.10}{64.00}{146}
\emline{72.10}{64.00}{147}{72.10}{64.00}{148}
\emline{72.10}{64.00}{149}{72.10}{63.00}{150}
\emline{72.10}{63.00}{151}{78.10}{63.00}{152}
\emline{78.10}{63.00}{153}{78.10}{57.00}{154}
\emline{78.10}{57.00}{155}{83.10}{57.00}{156}
\emline{83.10}{57.00}{157}{83.10}{51.00}{158}
\emline{83.10}{51.00}{159}{86.10}{51.00}{160}
\emline{58.10}{73.00}{161}{58.10}{76.00}{162}
\emline{60.10}{75.00}{163}{60.10}{76.00}{164}
\emline{82.10}{47.00}{165}{83.10}{47.00}{166}
\emline{39.10}{76.00}{167}{44.10}{71.00}{168}
\emline{44.10}{71.00}{169}{39.10}{66.00}{170}
\emline{39.10}{66.00}{171}{44.10}{61.00}{172}
\emline{44.10}{61.00}{173}{39.10}{56.00}{174}
\emline{39.10}{56.00}{175}{44.10}{51.00}{176}
\emline{44.10}{51.00}{177}{39.10}{46.00}{178}
\emline{39.10}{46.00}{179}{44.10}{41.00}{180}
\emline{44.10}{41.00}{181}{39.10}{36.00}{182}
\emline{39.10}{36.00}{183}{44.10}{31.00}{184}
\emline{44.10}{31.00}{185}{39.10}{26.00}{186}
\emline{39.10}{26.00}{187}{44.10}{21.00}{188}
\emline{44.10}{21.00}{189}{39.10}{16.00}{190}
\emline{39.10}{16.00}{191}{44.10}{11.00}{192}
\emline{44.10}{11.00}{193}{43.10}{10.00}{194}
\emline{37.10}{76.00}{195}{37.10}{76.00}{196}
\emline{35.10}{76.00}{197}{40.10}{71.00}{198}
\emline{40.10}{71.00}{199}{35.10}{66.00}{200}
\emline{35.10}{66.00}{201}{40.10}{61.00}{202}
\emline{40.10}{61.00}{203}{35.10}{56.00}{204}
\emline{35.10}{56.00}{205}{40.10}{51.00}{206}
\emline{40.10}{51.00}{207}{35.10}{46.00}{208}
\emline{35.10}{46.00}{209}{40.10}{41.00}{210}
\emline{40.10}{41.00}{211}{35.10}{36.00}{212}
\emline{35.10}{36.00}{213}{40.10}{31.00}{214}
\emline{40.10}{31.00}{215}{35.10}{26.00}{216}
\emline{35.10}{26.00}{217}{40.10}{21.00}{218}
\emline{40.10}{21.00}{219}{35.10}{16.00}{220}
\emline{35.10}{16.00}{221}{40.10}{11.00}{222}
\emline{40.10}{11.00}{223}{39.10}{10.00}{224}
\emline{46.10}{76.00}{225}{46.10}{76.00}{226}
\emline{44.10}{76.00}{227}{49.10}{71.00}{228}
\emline{49.10}{71.00}{229}{44.10}{66.00}{230}
\emline{44.10}{66.00}{231}{49.10}{61.00}{232}
\emline{49.10}{61.00}{233}{44.10}{56.00}{234}
\emline{44.10}{56.00}{235}{49.10}{51.00}{236}
\emline{49.10}{51.00}{237}{44.10}{46.00}{238}
\emline{44.10}{46.00}{239}{49.10}{41.00}{240}
\emline{49.10}{41.00}{241}{44.10}{36.00}{242}
\emline{44.10}{36.00}{243}{49.10}{31.00}{244}
\emline{49.10}{31.00}{245}{44.10}{26.00}{246}
\emline{44.10}{26.00}{247}{49.10}{21.00}{248}
\emline{49.10}{21.00}{249}{44.10}{16.00}{250}
\emline{44.10}{16.00}{251}{49.10}{11.00}{252}
\emline{49.10}{11.00}{253}{48.10}{10.00}{254}
\emline{132.10}{76.00}{255}{137.10}{71.00}{256}
\emline{137.10}{71.00}{257}{132.10}{66.00}{258}
\emline{132.10}{66.00}{259}{137.10}{61.00}{260}
\emline{137.10}{61.00}{261}{132.10}{56.00}{262}
\emline{132.10}{56.00}{263}{137.10}{51.00}{264}
\emline{137.10}{51.00}{265}{132.10}{46.00}{266}
\emline{132.10}{46.00}{267}{137.10}{41.00}{268}
\emline{137.10}{41.00}{269}{132.10}{36.00}{270}
\emline{132.10}{36.00}{271}{137.10}{31.00}{272}
\emline{137.10}{31.00}{273}{132.10}{26.00}{274}
\emline{132.10}{26.00}{275}{137.10}{21.00}{276}
\emline{137.10}{21.00}{277}{132.10}{16.00}{278}
\emline{132.10}{16.00}{279}{137.10}{11.00}{280}
\emline{137.10}{11.00}{281}{136.10}{10.00}{282}
\emline{130.10}{76.00}{283}{130.10}{76.00}{284}
\emline{141.10}{76.00}{285}{146.10}{71.00}{286}
\emline{146.10}{71.00}{287}{141.10}{66.00}{288}
\emline{141.10}{66.00}{289}{146.10}{61.00}{290}
\emline{146.10}{61.00}{291}{141.10}{56.00}{292}
\emline{141.10}{56.00}{293}{146.10}{51.00}{294}
\emline{146.10}{51.00}{295}{141.10}{46.00}{296}
\emline{141.10}{46.00}{297}{146.10}{41.00}{298}
\emline{146.10}{41.00}{299}{141.10}{36.00}{300}
\emline{141.10}{36.00}{301}{146.10}{31.00}{302}
\emline{146.10}{31.00}{303}{141.10}{26.00}{304}
\emline{141.10}{26.00}{305}{146.10}{21.00}{306}
\emline{146.10}{21.00}{307}{141.10}{16.00}{308}
\emline{141.10}{16.00}{309}{146.10}{11.00}{310}
\emline{146.10}{11.00}{311}{145.10}{10.00}{312}
\emline{139.10}{76.00}{313}{139.10}{76.00}{314}
\emline{137.10}{76.00}{315}{142.10}{71.00}{316}
\emline{142.10}{71.00}{317}{137.10}{66.00}{318}
\emline{137.10}{66.00}{319}{142.10}{61.00}{320}
\emline{142.10}{61.00}{321}{137.10}{56.00}{322}
\emline{137.10}{56.00}{323}{142.10}{51.00}{324}
\emline{142.10}{51.00}{325}{137.10}{46.00}{326}
\emline{137.10}{46.00}{327}{142.10}{41.00}{328}
\emline{142.10}{41.00}{329}{137.10}{36.00}{330}
\emline{137.10}{36.00}{331}{142.10}{31.00}{332}
\emline{142.10}{31.00}{333}{137.10}{26.00}{334}
\emline{137.10}{26.00}{335}{142.10}{21.00}{336}
\emline{142.10}{21.00}{337}{137.10}{16.00}{338}
\emline{137.10}{16.00}{339}{142.10}{11.00}{340}
\emline{142.10}{11.00}{341}{141.10}{10.00}{342}
\emline{98.10}{49.00}{343}{98.10}{52.00}{344}
\emline{98.10}{52.00}{345}{104.10}{52.00}{346}
\emline{104.10}{52.00}{347}{104.10}{58.00}{348}
\emline{104.10}{58.00}{349}{110.10}{58.00}{350}
\emline{110.10}{58.00}{351}{110.10}{64.00}{352}
\emline{110.10}{64.00}{353}{115.10}{64.00}{354}
\emline{115.10}{64.00}{355}{115.10}{64.00}{356}
\emline{115.10}{64.00}{357}{116.10}{64.00}{358}
\emline{116.10}{64.00}{359}{116.10}{70.00}{360}
\emline{116.10}{70.00}{361}{122.10}{70.00}{362}
\emline{122.10}{70.00}{363}{122.10}{75.00}{364}
\emline{96.10}{51.00}{365}{96.10}{54.00}{366}
\emline{96.10}{54.00}{367}{102.10}{54.00}{368}
\emline{102.10}{54.00}{369}{102.10}{60.00}{370}
\emline{102.10}{60.00}{371}{108.10}{60.00}{372}
\emline{108.10}{60.00}{373}{108.10}{66.00}{374}
\emline{108.10}{66.00}{375}{113.10}{66.00}{376}
\emline{113.10}{66.00}{377}{113.10}{66.00}{378}
\emline{113.10}{66.00}{379}{114.10}{66.00}{380}
\emline{114.10}{66.00}{381}{114.10}{72.00}{382}
\emline{114.10}{72.00}{383}{120.10}{72.00}{384}
\emline{120.10}{72.00}{385}{120.10}{77.00}{386}
\emline{94.10}{53.00}{387}{94.10}{56.00}{388}
\emline{94.10}{56.00}{389}{100.10}{56.00}{390}
\emline{100.10}{56.00}{391}{100.10}{62.00}{392}
\emline{100.10}{62.00}{393}{106.10}{62.00}{394}
\emline{106.10}{62.00}{395}{106.10}{68.00}{396}
\emline{106.10}{68.00}{397}{111.10}{68.00}{398}
\emline{111.10}{68.00}{399}{111.10}{68.00}{400}
\emline{111.10}{68.00}{401}{112.10}{68.00}{402}
\emline{112.10}{68.00}{403}{112.10}{74.00}{404}
\emline{112.10}{74.00}{405}{118.10}{74.00}{406}
\emline{118.10}{74.00}{407}{118.10}{79.00}{408}
\emline{94.10}{53.00}{409}{94.10}{51.00}{410}
\emline{96.10}{51.00}{411}{96.10}{49.00}{412}
\emline{98.10}{49.00}{413}{98.10}{47.00}{414}
\emline{98.10}{47.00}{415}{97.10}{47.00}{416}
\emline{122.10}{75.00}{417}{122.10}{76.00}{418}
\put(108.24,00.00){\makebox(0,0)[cc]{Fig.3. Diagram with the eikonalized
pomeron exchanged }}
\emline{31.10}{76.00}{419}{23.10}{76.00}{420}
\emline{31.10}{10.00}{421}{23.10}{10.00}{422}
\emline{151.10}{76.00}{423}{161.10}{76.00}{424}
\emline{151.10}{10.00}{425}{161.10}{10.00}{426}
\emline{90.00}{38.00}{427}{90.00}{36.00}{428}
\emline{90.00}{36.00}{429}{93.00}{33.00}{430}
\emline{93.00}{33.00}{431}{89.00}{29.00}{432}
\emline{89.00}{29.00}{433}{93.00}{25.00}{434}
\emline{93.00}{25.00}{435}{89.00}{21.00}{436}
\emline{89.00}{21.00}{437}{93.00}{17.00}{438}
\emline{93.00}{17.00}{439}{89.00}{13.00}{440}
\emline{89.00}{13.00}{441}{92.00}{10.00}{442}
\end{picture}
\end{center}
\vskip 0.5cm
Evidently it is impossible to calculate such diagrams in a general form
without any simplified assumptions. At the same time there are two
important and interesting points. Firstly, does the violation of
unitarity bound indeed removed after an eikonalization? Secondly, how fast
the eikonalized diffractive cross-section $\sigma^{SD}$ rises at
$s\rightarrow \infty$?

Our model we define in the form, which corresponds to Fig.3.
$$M^{2}\frac{d\sigma^{SD}}{dM^{2}} = \sigma_{0}^{-1} \tilde G\,I,$$
where
$$I = \int
d\vec{b}d\vec{b'}\exp(-2\Omega(s,b))\, \biggl\{1-\exp[-\tilde
\Omega(s_{0}\frac{s}{M^{2}},b')]\biggr\}^{2}\, \biggl\{1-\exp[-\tilde
\Omega(M^{2},|\vec{b}-\vec{b'}|)]\biggr\}\eqno(10)$$
and $\tilde G$ includes all of the relevant couplings and
constants.  The value $\tilde \Omega(z,b)$ differs of the input
one-pomeron $\Omega(z,b)$, (exp.  (4)). The difference is that one of
vertices $g$ (in $s,t$-representation) is changed for the part of
triple-pomeron vertex.
$$V_{PPP}(t_0;t_1,t_2)=v\exp(r_0^2t_0+r^2_{1,2}(t_1+t_2)),$$
specifically
$$g(t_i)\rightarrow v^{1/3}\exp(r_i^2t_i).$$

We note that it is not enough to eikonalize only pomerons in $3P$-vertex
in order to restore unitarity. If we do that, the factor
$\exp(-2\Omega(s,b))$ is absent in the integrand of (10). Integral is
calculated easily
$$M^{2}\frac{d\sigma^{SD}}{dM^{2}} \propto
R^{2}_{1}R^{2}_{2}\ln(s/M^{2})\ln(M^{2}/s_{0}),$$
$$\sigma^{SD} \propto \ln^{5}(s/s_{0}).$$
The result evidently contradicts to the Froissart-Martin bound.

Let's now calculate the asymptotic behaviour of integral $I$ at the
limit $s\rightarrow \infty.$ To do that let's rewrite it
in the following form
\begin{eqnarray}
I=
4\pi\int\limits_{0}^{\infty}db\int\limits_{0}^{\infty}db'
\int\limits_{0}^{\pi}d\phi\,bb'\biggl[1-e^{-\nu_{1}e^{-b'^{2}/R_{1}^{2}}}
\biggr]
\biggl[1-e^{-\nu_{2}e^{-b^{2}/R_{2}^{2}}}
\biggr]^{2}
\nonumber\\
\times \exp \biggl\{-2\nu_{0}\exp[-(b^{2}+b'^{2}+2bb'\cos
\phi)/R^{2}_{0}]\biggr\}.
\nonumber
\end{eqnarray}
To avoid the extra cumbersations in the following formulae we use
the notations:
$$\nu_{0}\equiv \nu(s/s_{0}),\quad \nu_{1}=
\frac{4gv}{R^{2}_{1}}\biggl(\frac{M^{2}}{s_{0}}\biggr)^{\Delta},\quad
\nu_{2}=
\frac{4gv}{R^{2}_{2}}\biggl(\frac{s}{M^{2}}\biggr)^{\Delta},\quad $$
$$R^{2}_{0}\equiv R^{2}(s/s_{0}),\quad R^{2}_{1}\equiv
\tilde R^{2}_{1}(M^{2}/s_{0}),\quad R^{2}_{2}\equiv \tilde
R^{2}_{2}(s/M^{2}),$$
where
$R^{2}(s/s_{0})$ is defined by exp.(5), and
$$\tilde
R^{2}_{1}(M^{2}/s_{0})=B_{0}+r^{2}_{0}+4\alpha'\ln(M^{2}/s_{0}),\qquad
\tilde R^{2}_{2}(s/M^{2})=B_{0}+r^{2}_{1,2}+4\alpha'\ln(s/M^{2}).$$
Making the substitution of the integration variables
$$\nu_{1}\exp
(-b^{2}/R_{1}^{2}) = z_{1}, \quad \nu_{2}\exp (-b^{2}/R_{2}^{2}) =
z_{2}$$
we get
\begin{eqnarray}
I & = & 2\pi R^{2}_{1}R^{2}_{2}
\int\limits_{0}^{\nu_{1}}\frac{dz_{1}}{z_{1}}
\int\limits_{0}^{\nu_{2}}\frac{dz_{2}}{z_{2}} \Bigl
(1-e^{-z_{1}}\Bigr)^{2}(1-e^{-z_{2}})\nonumber\\
& & \times\int\limits_{0}^{\pi}d\phi\,\exp \biggl\{
-2\nu_{0}\exp[-(\rho_{1}\ln( \frac{\nu_{1}}{z_{1}})+2
\sqrt{\rho_{1}\ln( \frac{\nu_{1}}{z_{1}})\rho_{2}\ln(
\frac{\nu_{2}}{z_{2}}) }\cos
\phi+\rho_{2}\ln(\frac{\nu_{2}}{z_{2}}))] \biggr\},\nonumber
\end{eqnarray}
where
$$\rho_{1}=\tilde R^{2}_{1}/R^{2}_{0}, \quad \rho_{2}=\tilde
R^{2}_{2}/R^{2}_{0}.$$

It is convenient to divide both integrals over
$z_{1}$ and $z_{2}$ into two parts: one is from $0$ up to
$1$ and second is from $1$ up to $\nu_{i}$.
$$I =
I^{1,1}_{0,0}+I^{\nu_{1},1}_{1,0}+I^{1,\nu_{2}}_{0,1}
+I^{\nu_{1},\nu_{2}}_{1,1}.$$
Consider for example first of them  where the integrations over z's go
from $0$ up to $1$ (remaining ones are similarly estimated). The small
$z_{1}$ and $z_{2}$ contribute to it. Therefore replacing $1-\exp(-z_{i})$
by $z_{i}$ and using the new variables
$$x_{i} = - \frac{\ln z_{i}}{\ln
\nu_{i}},\quad i=1,2,$$
we write $I^{1,1}_{0,0}$ in the form
\begin{eqnarray}
I^{1,1}_{0,0} & = & 2\pi R^{2}_{1}R^{2}_{2}
\ln \nu_{1}\ln \nu_{2} \int\limits_{0}^{\pi}d\phi
\int\limits_{0}^{\infty} dx_{1}\exp(-2x_{1}\ln\nu_{1})
\int\limits_{0}^{\infty} dx_{2}\exp(-x_{2}\ln\nu_{2})\nonumber\\
& & \times \exp \biggl \{ -2\nu_{0}\exp [-(\beta_{1}\sqrt{1+x_{1}}
+\beta_{2}\sqrt{1+x_{2}})^2+4\beta_{1}\beta_{2}\sqrt{(1+x_{1})(1+x_{2})}\sin^{2}
\frac{\phi}{2} ] \biggr \},\nonumber
\end{eqnarray}
where
$$\beta_{1} = \sqrt{\rho_{1}\ln \nu_{1}},\quad
\beta_{2} = \sqrt{\rho_{2}\ln \nu_{2}}.\eqno(11)$$

Before the further calculations let's pay attention to the behaviour of
the $\rho,\, \beta,\, \nu$ at $s\rightarrow \infty$. It is easy to see
that for an arbitrary but fixed $\rho_{i}$
$$\ln\nu_{i}=\rho_{i}\ln\nu_{0}\cdot\{1-\frac{\ln(R^{2}_{0}/\sigma_{0})}
{\ln\nu_{0}}+o(\frac{1}{\ln(s/s_{0})})\},\eqno(12a)$$
$$\beta_{i}=\rho_{i}\sqrt{\ln\nu_{0}}\cdot\{1-\frac{1}{2}
\frac{\ln(R^{2}_{0}/\sigma_{0})}{\ln\nu_{0}}+
o(\frac{1}{\ln(s/s_{0})})\},\eqno(12b)$$
$$\rho_{1}+\rho_{2}=1+\frac{r^{2}_{0}+r^{2}_{1,2}}{R^{2}_{0}},\quad
(\beta_{1}+\beta_{2})^{2} = \ln\nu_{0} -
\ln \biggl( \frac{R_{1}^{2}R_{2}^{2}}{\sigma_{0}R^{2}_{0}} \biggr) +
o(1).\eqno(12c)$$
Making use of above properties of $\rho,\, \beta,\, \nu,$ one can
argue that the main contribution in the integrals over $x_{i},\phi$
is determined by small $x_{1},$\,$x_{2},$\,$\phi$.
Keeping the linear in $x_{1},$\,$x_{2},$\,$\phi^{2}$ terms in the
internal exponential and substituting
$$u_{1} = \exp(-\beta_{1}(\beta_{1}+\beta_{2})x_{1}),\quad
u_{2} = \exp(-\beta_{2}(\beta_{1}+\beta_{2})x_{2}),$$
we obtain
\begin{eqnarray}
I^{1,1}_{0,0} & = &2\pi \frac{R^{2}_{1}R^{2}_{2}
\ln\nu_{1}\ln\nu_{2}}{\beta_{1}\beta_{2}(\beta_{1}+\beta_{2})^{2}}
\int\limits_{0}^{\pi}d\phi\int\limits_{0}^{1}\frac{du_{1}}{u_{1}}
\int\limits_{0}^{1}\frac{du_{2}}{u_{2}}\,u_{1}^{a_{1}}u_{2}^{a_{2}}
\exp\biggl\{-2\frac{R_{1}^{2}R_{2}^{2}}{\sigma_{0}R^{2}_{0}}
\exp(\beta_{1}\beta_{2}\phi^{2})u_{1}u_{2}\biggr\}\simeq \nonumber \\
& & \nonumber\\
& &\nonumber \\
& & 2\pi R^{2}_{1}R^{2}_{2}
\frac{\sqrt{\pi}\ln\nu_{1}\ln\nu_{2}}{(\beta_{1}\beta_{2})^{3/2}
(\beta_{1}+\beta_{2})^{2}(a_{1}-a_{2})}
\biggl\{\frac{\Gamma(a_{2})}{\sqrt{a_{2}}}
\biggl(2\frac{R_{1}^{2}R_{2}^{2}}{\sigma_{0}R^{2}_{0}}\biggr)^{-a_{2}} -
\frac{\Gamma(a_{1})}{\sqrt{a_{1}}}
\biggl(2\frac{R_{1}^{2}R_{2}^{2}}{\sigma_{0}R^{2}_{0}}\biggr)^{-a_{1}}
\biggr\}, \nonumber
\end{eqnarray}
where
$$a_{1} = \frac{\beta_{1}/\rho_{1}}{\beta_{1}+\beta_{2}},\quad
a_{2} = \frac{2\beta_{2}/\rho_{2}}{\beta_{1}+\beta_{2}}.\eqno(13)$$
Since $a_{1}=1+o(1/\ln(s/s_{0}))$ and
$a_{2}=2+o(1/\ln(s/s_{0}))$, as it follows from
(12),(13), we finally obtain for integral $I^{1,1}_{0,0}$
$$I^{1,1}_{0,0} = \pi \sqrt{\pi} \sigma_{0}R^{2}_{0}
\frac{\ln\nu_{1}\ln\nu_{2}}{(\beta_{1}\beta_{2})^{3/2}
(\beta_{1}+\beta_{2})^{2}}\biggl(1+o(\frac{1}{\ln(s/s_{0})})\biggr).$$

The similar calculations for another terms of integral $I$ give rise to
the following results
$$
I^{1,\nu_{2}}_{0,1} \simeq I^{1,1}_{0,0},\quad
I^{\nu_{1},1}_{1,0} \simeq
\frac{1}{2\sqrt{2}}\cdot\frac{\sigma_{0}R^{2}_{0}}{R^{2}_{1}R^{2}_{2}}
I^{1,1}_{0,0},\quad
I^{\nu_{1},\nu_{2}}_{1,1} <  const
\exp(-R^{2}_{0}/\sigma_{0}).$$

Thus at $s\rightarrow \infty$
$$M^{2}\frac{d\sigma^{SD}}{dM^{2}} \simeq
\pi \sqrt{\pi }\tilde G (R^{2}_{0})^{2}
\sqrt{\frac{\Delta\ln(s/s_{0})}{R^{2}_{1}R^{2}_{2}}} \simeq
const\frac{[\ln(s/s_{0})]^{3/2}}
{\sqrt{\ln(s/M^{2})\ln(M^{2}/s_{0})}}.\eqno(14)$$
Integration of the exp.(14) over $M^{2}$ in the domain where
$\rho_{1,2}\not \rightarrow 0$, gives
$$\sigma^{SD} \propto
\ln^{3/2}(s/s_{0}),\quad \sigma^{SD}/\sigma_{tot} \rightarrow 0 \quad
\mbox{ at } \quad s\rightarrow \infty.$$

Thus we have proposed and investigated the simplified eikonal model for
a process of hadronic single diffractive dissociation. We have shown
that eikonalization of each pomeron in the $3P$-diagram and account of the
elastic interaction of hadrons in the initial state allow to restore the
unitarity which is violated by an input supercritical pomeron. Our result is
only asymptotical one. The numerical calculations and account of the
nonasymptotical contributions are needed to compare the considered model
with experiment.

We thank very much A.I.Bugrij and L.L.Jenkovszky for the useful discussions
on the considered problem.

The research was made possible in part by Grant K4N100 from the Joint Fund
of the Government of Ukraine and International Science Foundation.
\vskip 1. cm
\begin{center}
{\bf References}
\end{center}
\begin{enumerate}
\item
N.Amos et al., Phys. Lett. {\bf B301}, 313 (1993);\\
CDF Collaboration: F.Abe et al., Phys. Rev. {\bf D50}, 5535 (1994).
\item
ZEUS Collaboration: M.Derric et al., Phys. Lett. {\bf B338}, 483 (1994);\\
H1 Collaboration: T.Ahmed et al., Phys. Lett. {\bf B348}, 681 (1995).
\item
A.Donnachie and P.Lahdshoff, Nucl. Phys. {\bf 231}, 189 (1984); Phys. Lett.
{\bf B296}, 227, 1992.
\item
T.T.Chou and C.N.Yang, Phys. Rev. {\bf 170}, 1591 (1968).
\item
O.A.Khrustaliov, V.E.Savrin and N.E.Tyurin, Elementary Particles and
Atomic Nuclear {\bf 7}, 21 (1976).
\item
E.M.Levin, Phys. Rev. {\bf D49}, 4469 (1994).
\item
E.Gotsman, E.M.Levin and U.Maor, Phys.Rev. {\bf D49}, 4321 (1994).
\item
A.Capella, J.Kaplan and J.Tran Thanh Van, Nucl. Phys. {\bf
B105}, 333 (1976).
\end{enumerate}
\end{document}